\documentclass[apj]{emulateapj}
\usepackage{natbib}
\usepackage{graphicx} 

\usepackage[normalem]{ulem}

\newcommand{\pder}[2]{\frac{\partial#1}{\partial#2}}
\newcommand{\Kappa}{K}

\begin{document}

\title{Energetic particle cross-field propagation early in a solar event}
\shorttitle{Early perpendicular propagation of SEPs}
\shortauthors{Laitinen et al.}

\author{T. Laitinen, S. Dalla, and M.S. Marsh}
\affil{Jeremiah Horrocks Institute, University of Central Lancashire,
  PR1 2HE Preston, UK}

\begin{abstract}
  Solar energetic particles (SEPs) have been observed to easily spread
  across heliographic longitudes, and the mechanisms responsible for
  this behaviour remain unclear. We use full-orbit simulations of a 10
  MeV proton beam in a turbulent magnetic field to study to what
  extent the spread across the mean field can be described as
  diffusion early in a particle event. We compare the full-orbit code
  results to solutions of a Fokker-Planck equation including spatial
  and pitch angle diffusion, and of one including also propagation of
  the particles along random-walking magnetic field lines. We find
  that propagation of the particles along meandering field lines is
  the key process determining their cross-field spread at 1 AU at the
  beginning of the simulated event. The mean square displacement of
  the particles an hour after injection is an order of magnitude
  larger than that given by the diffusion model, indicating that
  models employing spatial cross-field diffusion cannot be used
    to describe early evolution of an SEP event. On the other hand,
  the diffusion of the particles from their initial field lines is
  negligible during the first 5 hours, which is consistent with the
  observations of SEP intensity dropouts. We conclude that modelling
  SEP events must take into account the particle propagation along
  meandering field lines for the first 20 hours of the event.
\end{abstract}


\section{Introduction}\label{sec:introduction}

Solar Energetic Particles (SEPs), accelerated during solar eruptive
events, have been observed to have access to a wide range of
heliographic longitudes, both for impulsive \citep{Wiedenbeck2013} and
gradual \citep{Dresing2012} SEP events.  In a number of
  studies, the spreading of SEPs across the field has been modelled
  using the Fokker-Planck (FP) equation for the particle distribution
  function \citep[e.g.,][]{Jokipii1966} with field-aligned propagation
  implemented as diffusion in velocity space, and cross-field
  propagation as spatial diffusion across the mean field
  \citep[e.g.,][]{Zhang2009,Droge2010,He2011}. However, the
  cross-field diffusion coefficient $\kappa_\perp$ required to explain
  the SEP observations \citep[e.g.,][]{Zhang2003,Dresing2012}, is much
  larger than that derived from galactic cosmic ray observations
  \citep[e.g.][]{Burger2000} and full-orbit simulations
  \citep[e.g.][]{GiaJok1999}.

\citet[][]{Matthaeus2003} used a model where the particles
  diffuse along field lines that random-walk across the mean magnetic
  field direction, to study the particle propagation across the mean
  magnetic field. They obtained $\kappa_\perp$ that is consistent with
  the galactic cosmic ray observations and full-orbit simulations. The
  diffusive behaviour of particles in this model is an asymptotic,
  long-time solution.

In this paper we study whether the need for a large
$\kappa_\perp$ in the FP modelling of an SEP event is
due to the fact that its description of cross-field diffusion may not be
valid in the early phases of an SEP event. SEPs are observed at 1~AU
soon after their injection, at only a few scattering mean free paths
from their source \citep[e.g.][]{Palmer1982}. Thus the cross field
spreading may not have settled to the asymptotic diffusive behaviour
described by \cite{Matthaeus2003}. Therefore, the question arises of
whether a diffusion description for the initial SEP propagation is
appropriate.

We study the early time cross-field transport of an SEP event by means
of full-orbit simulations in a prescribed turbulence. We evaluate the
early cross-field transport in this model and compare it
quantitatively with that obtained from solution of a FP model.

Full-orbit models have been used to study charged particle propagation
in turbulence superposed onto a constant background field to study the
evolution and asymptotic values of the diffusion coefficients
\citep[e.g.][]{GiaJok1999,Qin2002,Qin2002apjl,QinEa2002,LaEa2012,LaEa2013a},
however not addressing the evolution of particle intensities at a
fixed location. \cite{GiaJokMaz2000} used a turbulence model in Parker
spiral geometry to study the effect of the size of the SEP source
region on the intensities observed at 1~AU, concluding that a small
source region would result in intensity dropouts such as observed by
\citet[e.g.,][]{Mazur2000}. In this study, however, we aim to quantify
the efficiency of the particle cross-field transport in a statistical
sense, and thus use a large source region.

To study the propagation early in an event, for simplicity we
superimpose turbulence on a constant background magnetic field. We
inject particles into the simulation as a beam, with pitch angle
cosine $\mu=1$, to mimic the initial strong focusing of
  particles in the radial field close to the sun, as this mechanism is
  absent in constant magnetic field. We compare the result of this
full-orbit simulation to a solution of a FP equation, and
quantitatively show that the early evolution of an SEP event cannot be
described using spatial cross-field diffusion in a FP equation.  The
early particle transport in the full-orbit simulations is consistent
with particles propagating along meandering field lines. We
demonstrate this by using a model that incorporates field line
meandering into the FP method.

\section{Models}\label{sec:model}

\subsection{Full-orbit simulations}

The full-orbit simulations are based on the description of the
turbulent magnetic field presented in \citet{GiaJok1999}, with
\begin{equation}\label{eq:turbfield}
\mathbf{B}(x,y,z)=B_0 \hat{\mathbf{z}}+\delta \mathbf{B}(x,y,z),
\end{equation}
where $B_0$ is a constant background field, along the $z$-axis, and $
\delta \mathbf{B}(x,y,z)$ a fluctuating field consisting of Fourier
modes. We use $B_0=5$~nT, consistent with the field strength at
  1 AU.  For the fluctuations, we use the composite model, where
turbulence is composed of slab and 2D components. We use a
  spectral index $\gamma=-1$ for the slab (1D) component, to prevent
  the so-called resonance gap issue which would complicate comparison
  with models using pitch angle diffusion, as discussed further in
  Section~\ref{sec:pitch-angle-spatial}. The amplitude of the
  turbulence is set to give a parallel mean free path of 0.3~AU for a
  10 MeV proton, a reasonable value for SEP protons
  \citep[e.g.,][]{Palmer1982}.  For the used spectral shape, the
  amplitude is somewhat lower than the interplanetary value, with
  parameter $B_1^2=0.1 B_0^2$ \citep[see][for definition]{LaEa2012}.

The full-orbit particle simulations follow the same approach as
\citet{LaEa2012}. We start the particles in a large volume, to exclude the effects of coherence by close-by fieldlines discussed by, e.g., \cite{GiaJokMaz2000} and
  \cite{Ruffolo2004}.  The quantities below are
  calculated relative to each particles' initial position so that in
  coordinates $x$, $y$ and $z$ the initial position of each particle
  is at the origin. Thus, our study models statistically the
    spreading of particles from a large source region, excluding the
    effects of local field line coherence.

We calculate the perpendicular variance of the particles,
$\sigma_i^2(z,t)=\left<(r_{\perp,i}(z,t)-\left<r_{\perp,i}(z,t)\right>)^2\right>$,
where $r_{\perp,i}=x,y$ and $\left<\right>$ represents the ensemble
average, as a function of time and location along the mean field. The
local running perpendicular diffusion coefficient, $\Kappa_{\perp
  i}(z,t)$, is defined as
\begin{equation}
  \label{eq:kappa}
  \Kappa_{\perp i}(z,t)=\frac{\sigma_i^2(z,t)}{2 t},
\end{equation}
and is obtained from particles within $z\pm\Delta z$, where $\Delta
z=15 r_\odot$, with $r_\odot$ the solar radius. These
  definitions of the perpendicular variance and local diffusion
  coefficient are not sensitive to the widening of the cross-field
  extent due to particles propagating along the mean field line:
  particles following their original field lines would produce a
  constant $\sigma_i^2(z,t)$, and any variation indicates decoupling
  of particles from the field lines
  \citep[e.g.][]{Hauff2010,Fraschetti2011perptimetheory,Fraschetti2012perptimesims}. Thus,
  $\sigma_i^2(z,t)$ is a powerful tool for determining the nature of
  the cross-field propagation of charged particles. It also better
  corresponds to what particle instruments observe: the
  intensities of particles at fixed locations, instead of the particle
  population's full spatial extent.

We also calculate the standard asymptotic values of the
 diffusion coefficients as
\begin{equation}
  \label{eq:asymptdiff}
  \kappa_i=\lim_{t\rightarrow\infty} \frac{\left<\xi(t)^2\right>}{2 t},
\end{equation}
where $\xi(t)=x,y,z$,  and the field line diffusion coefficient,
  due to the meandering of the turbulent field in
  Eq.~(\ref{eq:turbfield}), as
\begin{equation}
  \label{eq:asymptdiff_fl}
  D_{\perp,\xi}=\lim_{z\rightarrow\infty} \frac{\left<\xi(z)^2\right>}{2 z},
\end{equation}
where $\xi(z)=x(z),y(z)$ are the coordinates of the field line at mean
field direction distance $z$. These coefficients are used as
  input parameters in Sections~\ref{sec:pitch-angle-spatial}
  and~\ref{sec:FP-meandering}.

\subsection{Fokker-Planck test particle simulations}\label{sec:pitch-angle-spatial}

The second description of particle transport is based on a Fokker-Planck
equation appropriate for our model definitions of static turbulence on
constant background magnetic field, given by
\begin{equation}
  \label{eq:fokkerplanck}
  \pder{f}{t}+\mu
  v\pder{f}{z}=\pder{}{\mu}D_{\mu\mu}\pder{f}{\mu}+\nabla\cdot\left(\hat{\kappa}_\perp\nabla
  f \right),
\end{equation}
\citep[e.g.,][]{Jokipii1966,Schlickeiser2002crbook} where $f(x,y,z,v,\mu,t)$ is the particle
distribution function, $v$ and $\mu$ the particle's velocity and pitch
angle cosine, $D_{\mu\mu}$ the pitch angle diffusion coefficient, and 
\[
\hat{\kappa}_\perp=
\left( \begin{array}{ccc}
\kappa_\perp & 0 & 0 \\
0 & \kappa_\perp & 0 \\
0 & 0 & 0 \end{array} \right)
\] 
the cross-field spatial diffusion tensor. This equation is
solved via Monte Carlo test-particle simulations, using the
  approach described in e.g., \citet{Zhang2009} and
\citet{Droge2010}. Below, this model is referred to with abbreviation
FP.

The perpendicular diffusion of the particles, given by the last term
in Eq.~(\ref{eq:fokkerplanck}), is solved by means of stochastic
differential equations (SDE) \citep{Gardiner1985}, which gives
the perpendicular Monte Carlo step for the particles as
\begin{eqnarray}
  dx&=&\sqrt{2\kappa_\perp dt}\, W_{x} \label{eq:x_part} \\
  dy&=&\sqrt{2\kappa_\perp dt}\, W_{y} \label{eq:y_part}
\end{eqnarray}
where $W_{x}$ and $W_{y}$ are Gaussian random numbers with zero
mean and unit variance, and $dt$ is the time step length.

The parallel propagation of the particles is given by 
\begin{equation}
  dz=\mu\, v\, dt\label{eq:z_part}
\end{equation}
and the pitch angle is scattered isotropically using the method
introduced by \citet{Torsti1996}, with pitch angle diffusion
coefficient
\begin{equation}
  \label{eq:dmumu}
  D_{\mu\mu}=\nu\left(1-\mu^2\right),
\end{equation}
where the scattering frequency $\nu=v^2/(6\kappa_\parallel)$ is
  independent of pitch angle for turbulence spectral index $\gamma=-1$
  in quasilinear theory. This is chosen to avoid the problem of a
  resonance gap at small $\mu$ \citep[see,
  e.g.,][]{Schlickeiser2002crbook}, which would complicate comparison
  between the full-orbit simulations and the FP model.

We verified that the evolution of the pitch angle distribution
obtained by using the pitch-angle diffusion method of
\citet{Torsti1996} agrees well with that obtained from the full-orbit
simulations, and thus we are confident that the propagation along the
mean field lines is similar in the two methods. The two methods also
agree well with the analytical solution to the isotropic pitch angle
diffusion given by, e.g., \citet{Roelof1969}.

 \begin{figure*}
   \includegraphics[width=\textwidth]{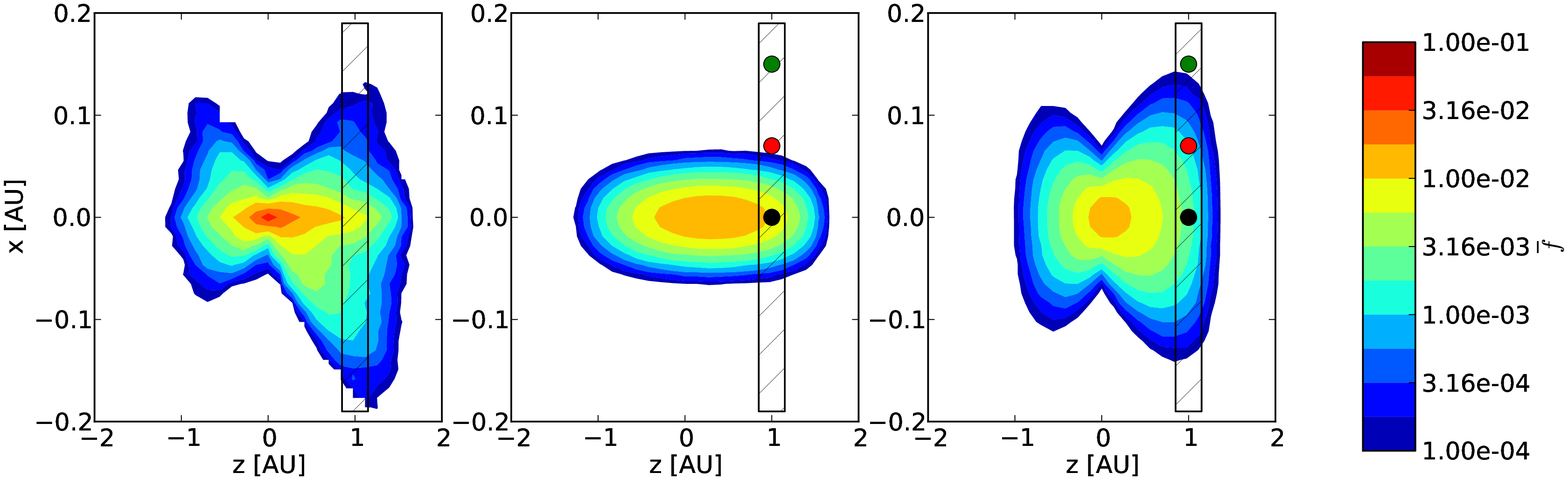}
   \caption{Contour plot of the spatial SEP distribution for the
     full-orbit (left panel), FP (centre panel) and
     FP+FLRW (right panel) models at 115 minutes from the
     injection. The box at $z=1\pm 0.15$~AU depicts the range that is used for calculating the variance and local running diffusion coefficients in Figs.~\ref{fig:diffcoeff} and ~\ref{fig:time_width}. The coloured circles correspond to the locations in Fig.~\ref{fig:time_int}.}\label{fig:contours}
 \end{figure*}

 \subsection{Fokker-Planck test particle simulations with meandering
   field lines}\label{sec:FP-meandering}

In the third model, we add the effect of meandering field lines to the
Fokker-Planck description of particle propagation introduced in
Section~\ref{sec:pitch-angle-spatial}. We model the field line
wandering as diffusion, using the field line diffusion coefficient
given by Eq.~(\ref{eq:asymptdiff_fl}). We solve the path of the fieldline using the SDE approach, which gives
\begin{eqnarray}
  dx_{B}&=&\sqrt{2 D_{\perp} dz_B}\, W_{x} \label{eq:x_field} \\
  dy_{B}&=&\sqrt{2 D_{\perp} dz_B}\, W_{y} \label{eq:y_field}.
\end{eqnarray}
We calculate the field line, $\left(x_B(z_B),y_B(z_B),z_B)\right)$
separately for each particle. The particle will then propagate along
this field line instead of the constant background field used in
Section~\ref{sec:pitch-angle-spatial}. 

The meandering of the field line is taken into account by advancing
the particle along the mean magnetic field direction by 
\[
dz=\mu\, v\, dt\, \cos\theta,
\]
where $\theta$ is the angle between the mean field and the local 
meandering field. The diffusion step, given by
Equations~(\ref{eq:x_part}) and~(\ref{eq:y_part}), is taken
perpendicular to the meandering field line. The particle's cross-field
deviation from its initial location is composed of the diffusive
propagation of the particle, $x(t)$ and $y(t)$, superimposed upon
the wandering of the field line described by, $x_{B}(z(t))$ and
$y_{B}(z(t))$. Below, this model will be referred to as FP+FLRW.

\section{Results and Discussion}\label{sec:results}

To analyse how energetic particles spread across the mean
magnetic field in the three propagation models described above, we
inject a population of 10 MeV protons with $\mu=1$ and follow them for
60 hours. The number of particles is $N=2\times 10^5$ in the
full-orbit simulations, while the FP and FP+FLRW simulations use
$N=2\times 10^6$ particles. The particle diffusion coefficients used in
the FP and FP+FLRW models are the
asymptotic values of the running diffusion coefficients of the
full-orbit simulations, as given by Eq.~(\ref{eq:asymptdiff}). For
this purpose, we simulate an isotropic proton population of N=2048
particles. For the turbulent field realisation presented in this
study, we obtain a parallel diffusion coefficient
$\kappa_\parallel=6.2\times 10^{21} \mathrm{cm}^2/\mathrm{s}$. The
perpendicular diffusion coefficient is $\kappa_\perp=6.6\times 10^{18}
\mathrm{cm}^2/\mathrm{s}$.

We also calculated the field line diffusion coefficient,
$D_\perp=2.1\times 10^{10} \mathrm{cm}$, from the magnetic field lines
used in the full-orbit simulations. This value is used to produce the
field line random walk, with
Eqs.~(\ref{eq:x_field})-(\ref{eq:y_field}).

We show the particle distribution with a contour plot in
Fig.~\ref{fig:contours}, for the full-orbit (left), FP (middle) and
FP+FLRW simulations (right panel). The contours represent the particle
distribution integrated along the $y$-direction,
$\overline{f}(x,z,t)$, 115 minutes after injection, with the
horizontal axis along the mean magnetic field. In the left panel
particles on both the negative and positive $z$ region expand in the
cross-field direction as they propagate farther from the origin along
the $z$-axis. The particles with $z<0$ have been scattered back from
the $z>0$ region, crossing the $z=0$ boundary close to
the origin. This pattern of propagation can be expected for
particles that follow field lines fanning out from
the origin, while decoupling from their initial field line with a slow
rate. The full-orbit particle distribution is distinctly different
from the elliptical profile obtained with the FP model (middle panel), but
qualitatively similar to the FP+FLRW model (right panel).

 \begin{figure}
 \includegraphics[width=\columnwidth]{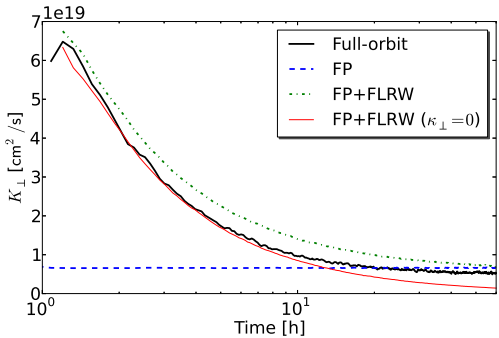}
   \caption{The local running diffusion coefficient for the models,
     determined from particles at 1 AU from the injection
     location (the boxes in Fig.~\ref{fig:contours}).\label{fig:diffcoeff}}
 \end{figure}

In Fig.~\ref{fig:diffcoeff} we show $\Kappa_\perp(z,t)$, as defined by Eq.~(\ref{eq:kappa}), calculated using the particles in the range depicted by the boxes in Fig.~\ref{fig:contours}.  The temporal evolution of
$\Kappa_{\perp}(z=1\mathrm{~AU},t)$ obtained with the FP model (blue dashed curve)
differs considerably from the other models, staying at a constant
value, as can be expected. In the full-orbit simulations (solid black
curve), at the time of arrival of the first particles at 1 AU, about an
hour after injection, $\Kappa_{\perp}$ is an order of magnitude
larger than the FP value. The full-orbit $\Kappa_{\perp}$
reaches the level of the FP description only 10 hours after injection.

\begin{figure}
 \includegraphics[width=\columnwidth]{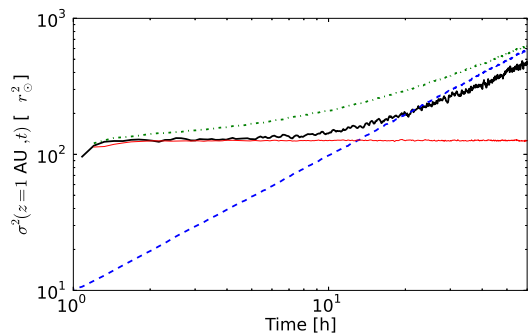}
 \caption{The perpendicular variance of particles observed at
   1~AU (the boxes in Fig.~\ref{fig:contours}). For legend, see Fig.~\ref{fig:diffcoeff}.\label{fig:time_width}}
 \end{figure}

For the first 5 hours of the simulated event, the full-orbit diffusion
coefficient follows closely the curve for a FP+FLRW model with
$\kappa_\perp=0$ (red curve), which describes particles remaining
on their original field lines. This can be seen more clearly in
Fig.~\ref{fig:time_width}, where we show the evolution of the cross-field variance at 1~AU, $\sigma_\perp^2(z=\mbox{1 AU},t)$. The
variance of the full-orbit simulated particles remains constant for
the first 5 hours from injection, indicating that the
FLRW effect dominates over particles decoupling from their field
lines. As discussed by \cite{Giajok2012}, the dominance of the
field line meandering over the decoupling can
explain the dropouts in SEP intensities observed in some impulsive
events \citep[e.g.,][]{Mazur2000,Chollet2011}. Only at later times
does the decoupling of the particles from their initial field lines
become non-negligible, and the full-orbit running diffusion
coefficient approaches the FP+FLRW with non-zero perpendicular
coefficient (the dash-dotted green curve in Figs.~\ref{fig:diffcoeff}
and \ref{fig:time_width}).

It should be noted that the non-monotonic behaviour and the
  small deviation from the red curve in the initial phase of the
full-orbit simulations is  caused by the
local structures in the particular turbulence realisation; this
behaviour varies between realisations.

 \begin{figure}
 \includegraphics[width=\columnwidth]{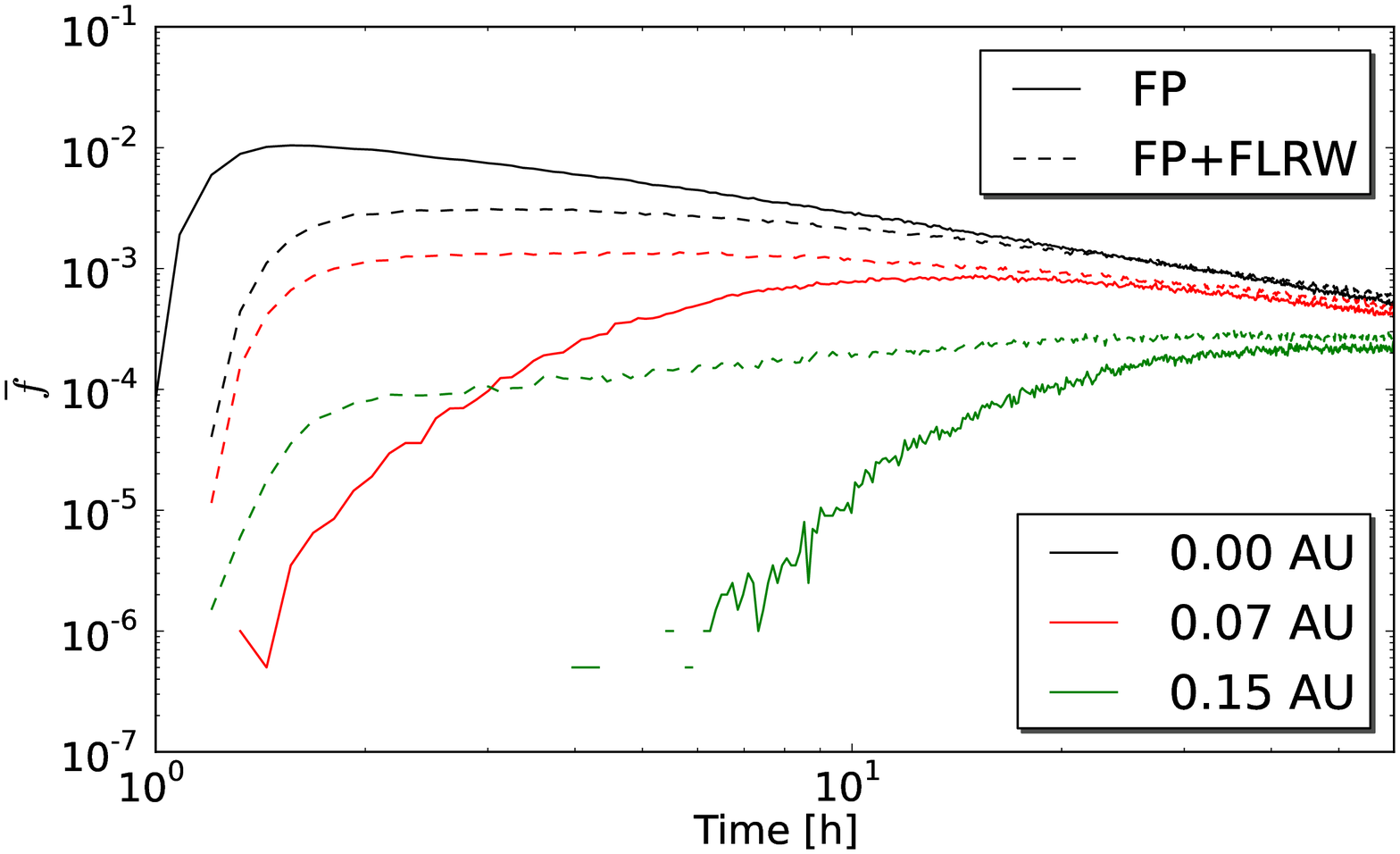}
 \caption{Proton distribution function $\overline{f}(x,z,t)$ at the locations marked by black, red and green circles in Fig.~\ref{fig:contours}, shown by black, red and green curves, respectively. \label{fig:time_int}}
 \end{figure}

In Fig.~\ref{fig:time_int}, we show the evolution of
$\overline{f}(x,z,t)$, at different cross-field locations $x$ at parallel
distance $z=1$~AU from the injection location. The
solid curves correspond to the FP model. We do not show the full-orbit
simulations, as the local structure and asymmetries, evident from
Fig.~\ref{fig:contours}, are complicated and vary between turbulence
realisations. Instead, we show the FP+FLRW model with non-zero
$\kappa_\perp$ (dashed curves), which, of the FP models presented in this paper, best
reproduces the evolution of the cross-field extent of the full-orbit
simulated particles at 1 AU (Fig.~\ref{fig:time_width}).

At cross-field location $x=0$~AU (black curves), corresponding to a
nominal connection to the injection site, the intensity evolution in
the two models is similar.  The FP+FLRW intensities are delayed by 15 minutes
  with respect to the FP model, due to the longer pathlength of the
  particle along the meandering field line. Full-orbit simulation
  onsets, not shown, variy somewhat with different realisations, due
  to varying field line lengths.

At a location not nominally connected to the injection site (red and
green curves), the FP and FP+FLRW models show very different
evolutions: while the FP+FLRW model shows rapid increase at
wide cross-field range, the FP model displays significant delay in
intensity onset, with a several-hour difference between the two
models.

Our results thus suggest that the early evolution across the mean
field is not diffusive, and using an incorrect model for the particle
propagation may significantly distort the interpretation of the observed SEP
intensities. The particles following the meandering field lines have
access to large cross-field distances, compared to diffusively
propagating particles, and arrive more promptly at the observing
spacecraft if it is not magnetically well connected to the
injection location. Using a spatial diffusion model to analyse such an
event may result in either a large cross-field diffusion coefficient,
or an interpretation of an extended acceleration region.

In this work, we introduced simplifications, such as the constant
background magnetic field, to be able to compare different approaches
with as few ambiguities as possible. Recent work by \citet{Giajok2012}
found that in a Parker spiral geometry, spatial diffusion would be
able to spread particles up to $180^\circ$ in longitude, and our
results suggest that adding the meandering of field lines as in this
study would aid the particles to fill the inner heliosphere with SEPs
even further. The recently discussed effects of large-scale drifts for
SEPs in Parker field may also increase the spread of SEPs in
interplanetary space \citep{Dalla2013,Marsh2013}. On the other hand,
the field line meandering may be limited by structures in the solar
wind, as discussed by \cite{LaEa2012}. Also the adiabatic focusing and
deceleration of the particles, occurring in the expanding corona, are
important for SEP transport
\cite[e.g.,][]{Ruffolo1995,Kocharov1998}. Furthermore, the
  radial and spectral evolution of heliospheric turbulence
  \citep[see, e.g.,][and references therein]{Cranmer2005} will affect
  the spreading of particles in interplanetary space.

We also note that the method of calculating the cross-field variance
at a fixed distance along the mean field from the
initial position, $\sigma^2(z,t)$, may be useful for studying the rate at
which particles decouple from the field lines. Decoupling can
be clearly seen at $z=0$ in the left panel of
Fig.~\ref{fig:contours}: the cross-field spreading around $x=0,z=0$ is caused by the decoupling, and the rate can be
measured using the approach introduced in this work. As can be seen in
Fig.~\ref{fig:diffcoeff}, a constant spatial diffusion
coefficient, as used in the FP+FLRW model, cannot describe this
spreading. It is likely that the spreading is connected to the
evolution of the field line separation \citep[see][]{Ruffolo2004}
after the particle has moved to a different field line.

\section{Conclusions}\label{sec:conclusions}

In this work, we have studied particle propagation across the mean
field as a function of time, after an impulsive, beam-like injection
into a turbulent magnetic field.  We compared the spatial evolution of
energetic particles using three simulation methods: full-orbit
particle simulations using synthetic meandering field lines,
spatial and pitch angle diffusion along constant background
field (FP), and diffusive propagation along stochastically spreading
field lines (FP+FLRW). Our main findings are as follows:
\begin{itemize}
\item The propagation of particles across the mean magnetic field in
  the early phase of an event is mainly due to the particles following
  meandering magnetic fields, resulting in cross-field mean square
  width an order of magnitude larger than that predicted by a diffusion
  model. This behaviour of the particle propagation cannot be
    described as a cross-field diffusion across mean magnetic
    field, but requires a description for the meandering field lines.
\item The large cross-field spread of the particles at the time of
  arrival of the first particles (Fig.~\ref{fig:time_width}) may
  explain the observations of SEPs at wide longitudinal separation, as
  reported by, e.g., \citet{Dresing2012} and \citet{Wiedenbeck2013}.
\item The timing of the access to field lines at large cross-field
  distances is rapid in a description including FLRW (red and green
  dashed curves in Fig.~\ref{fig:time_int}), and takes place much
  faster than in a model including diffusion only.
\end{itemize}

Early in an event the particles remain well on their initial field
lines, and thus do not propagate across the meandering field lines
(see Figs.~\ref{fig:diffcoeff} and~\ref{fig:time_int}). A vanishingly
small cross-field diffusion coefficient during impulsive ``dropout''
events has been previously suggested based on FP simulations
\citep{Droge2010} and observations \citep{Chollet2011}. We find, based
on our simulations, that a small diffusion coefficient early in an SEP
event is not contradicted by a larger diffusion coefficient later
in the event.

We conclude that in modelling of an SEP event, the description of
cross-field propagation as spatial diffusion only is not sufficient,
and a description of field line meandering should be used. Adding FLRW
to the standard FP description, as presented in this paper, is a
reasonable method to model the particle propagation as seen in the
full orbit simulations. However, the spatial cross-field diffusion
from the meandering field lines is very slow during the first hours of
the event, approaching the asymptotic diffusion coefficient only at
later times. Thus, a model with time-dependent diffusion coefficient
is needed to accurately describe the particle decoupling from their
initial field lines.

\acknowledgments

We acknowledge support from the UK Science and Technology Facilities
Council (STFC) (grant ST/J001341/1) and from the European Commission
FP7 Project COMESEP (263252). Access to the University of Central
Lancashire's High Performance Computing Facility is gratefully
acknowledged.

\end{document}